\documentclass[12pt,preprint]{aastex}
\begin{document}

\newcommand{\be}{\begin{eqnarray}}
\newcommand{\ee}{\end{eqnarray}}
\newcommand{\vperp}{V_{\perp}}

\title{PSR~J1740+1000: A Young Pulsar Well \\ Out Of The Galactic Plane}
\author{M.\ A.\ McLaughlin\altaffilmark{1}, Z.\ Arzoumanian\altaffilmark{2}, J.\ M.\ Cordes\altaffilmark{1}, D.\ C. Backer\altaffilmark{3}, \\ A.\ N.\ Lommen\altaffilmark{3}, D.\ R.\ Lorimer\altaffilmark{4}, \& A. F. Zepka\altaffilmark{5}}
\altaffiltext{1}{Astronomy Department, Cornell University, Ithaca, NY 14853}
\altaffiltext{2}{Laboratory for
        High-Energy Astrophysics, NASA-GSFC, Code 662, Greenbelt, MD
        20771. NAS/NRC Research Associate}
\altaffiltext{3}{Astronomy Department, University of California, Berkeley, CA 94720}
\altaffiltext{4}{Arecibo Observatory, HC3 Box 53995,
Puerto Rico 00612}
\altaffiltext{5}{Currently at Numerical Technologies, 70 W. Plumeria Drive, San Jose, CA 95134-2134}
\begin{abstract}

We discuss PSR~J1740+1000, one of five pulsars recently
discovered in a search of 470 square degrees at
430 MHz during the upgrade of the 305-m Arecibo telescope. 
The period $P = 154$ ms  and 
period derivative $\dot P = 2.1\times10^{-14}$ s s$^{-1}$ 
imply a spin-down age  $\tau_s = P/2\dot P = 114$ kyr that 
is smaller than 95\% of all known pulsars. The youth and proximity of this
pulsar make it a good candidate for detection at X-ray and gamma-ray energies. 
Its high Galactic latitude ($b = 20.4^{\circ}$)
suggests a very high velocity if the pulsar was born
in the midplane of the Galaxy and if its kinematic age equals its
spindown age.  Interstellar scintillations, however, suggest a much lower
velocity.   We discuss possible explanations
for this discrepancy, taking into account
(a) possible birth sites away from the midplane;
(b) contributions from the unmeasured radial velocity; (c) a kinematic age different from the spin-down age;
and (d) biasing of the scintillation velocity by enhanced scattering
from the North Polar Spur. 

\end{abstract}


\section{Introduction}

PSR~J1740+1000, discovered in an Arecibo survey at 430~MHz,
is a young pulsar at high Galactic latitude.
It may be representative of a class of high-velocity pulsars
which have thus far eluded detection. Determining the velocity and
origin of this object is important as
pulsar velocities represent fossil information about evolution
of close binary systems and core collapse processes in
supernovae.  The largest velocities, in excess of
1000 km s$^{-1}$, provide the greatest constraints on such processes.
Measuring pulsar velocities is also crucial for pulsar population statistics,
determining the birth rate and birth places of pulsars and for planning future 
pulsar searches.
Unfortunately, pulsar surveys have been strongly biased against the detection of
 high-velocity
pulsars. While the average pulsar speed is about 500 km s$^{-1}$, the tail of the velocity
distribution extends to at least 1600 km s$^{-1}$, with 10-20\% of all pulsars having
speeds greater than 1000 km s$^{-1}$ \cite{lyne94,cher98,arz01}.
At these speeds, within a typical 10 Myr pulsar lifetime many high-velocity pulsars will have
moved
beyond the
volume of detectability of previous surveys at low Galactic latitudes 
\cite{lor97}.
 
In \S 2 of this paper, we describe the search
and follow-up observations. In \S 3, we present a timing model and discuss the pulsar's
single pulse and interstellar scintillation (ISS) properties.   In \S 4, we explore ways to
reconcile the kinematic and ISS velocities of the pulsar. In \S 5 and \S 6, we describe prospects
for astrometry and high-energy detection, respectively. In \S 7, we present our main conclusions and 
plans for future study. 

\section{Observations}

PSR~J1740+1000 was discovered during the
upgrade of the Arecibo Observatory, when several collaborations
jointly surveyed the Arecibo sky (covering declinations roughly between $-1^{\rm o}$
and 39$^{\rm o}$). As the telescope was not able to track sources, data were taken 
as the sky drifted
through the beam at the sidereal rate. A point source drifts through the Arecibo
430-MHz 10$'$ beam in approximately 40 seconds. The feed was moved once a day to allow
successive declination strips to be surveyed. The nominal sensitivity of this search, which
employed 32 channels covering a bandwidth of 8 MHz, was
approximately 0.5 mJy. We estimate that the techniques used for radio frequency
interference excision and
candidate evaluation
raised our sensitivity to roughly 1 mJy. With a sampling interval of 250 $\mu$s,
 we were sensitive to pulsars with periods $\sim$ 1 ms up to dispersion measures (DMs)
of $\sim$ 20 pc cm$^{-3}$
and to pulsars with periods $\sim$ 30 ms up to DMs of $\sim$ 600 pc cm$^{-3}$.

 The Berkeley/Cornell collaboration covered 470 square degrees 
between 1994 October and 1995 February.
We detected 5 new pulsars in the search. Three
(J0030+0451, J0711+0931 and J1313+0931)
were reported in Lommen et al. (2000), and two (J1740+1000 and J1849+2423) were reported in
McLaughlin et al. (2000).  
With a search-derived period of 154 ms,
a DM of $\sim$ 20 pc cm$^{-3}$ and signal-to-noise ratio of  $\sim$ 20,
J1740+1000 was one of several pulsar candidates which were reobserved after the upgrade was completed.
Confirming observations were done in May 1998 with the Arecibo Observatory Fourier
Transform Machine (AOFTM), a Fourier transform-based spectrometer with 1024 10-kHz
frequency channels and 102.4 $\mu$s sampling\footnote{http://www.naic.edu/\~\,aoftm}.
The pulsar
was timed from May 1999 onwards with the Penn State
Pulsar Machine (PSPM), a filterbank
with 128 60-kHz frequency channels and 80 $\mu$s
sampling. We have also observed this pulsar with
the Wideband Arecibo Pulsar Processor (WAPP), a fast-dump digital correlator covering up to
a 100 MHz
bandwidth with a programmable number of lags and time
 resolution\footnote{http://www.naic.edu/\~\,wapp} \cite{dowd00}.

We mapped the
field of this pulsar with the Very Large Array (VLA) in A array at 1.4 GHz 
using a 50-MHz bandwidth and 12-minute integration. 
At the time of these observations in July 1999,
the pulsar's position was known to approximately $3'$ (i.e. one 1.4-GHz
beam at Arecibo). 
We detected
a continuum source in this observation with a flux
density  of 2.26 $\pm$ 0.17 mJy at a
position
(J2000 = MJD 51385)
of RA=$17^{\rm h}40^{\rm m}25^{\rm s}.958 \pm 0.003$, DEC=$+10^{\circ}00'05.91\pm 0.05$ (derived
using the \tt AIPS \rm task \tt JMFIT\rm). Subsequent timing confirmed that this source
was indeed the pulsar. The quoted 0.17 mJy error on the flux density does not account
for fluctuations due to interstellar scintillation, which are significant.
No other point sources were
within $3'$ of the position of PSR~J1740+1000, but 2 sources,
both listed in the NVSS catalog \cite{condon98}, 
were within $5'$. No filamentary or bow-shock structures
were visible in the radio maps.

\section{Analysis and Results}
\subsection{Pulsar Properties}

We began
regular timing observations of  
PSR~J1740+1000 in May 1999 at an observing frequency of 
430 MHz. 
After we established an initial timing model and more accurate position,
timing 
observations were carried out predominantly at 1.4 GHz,
where higher signal-to-noise profiles 
could be acquired.
Pulse times-of-arrival (TOAs) were calculated by cross correlating dedispersed
profiles with templates, using separate templates made from the addition
of many individual pulses at both 430 MHz and 1.4 GHz.
These TOAs were then
analyzed using the
TEMPO software package\footnote{http://pulsar.princeton.edu/tempo}
 \cite{tay89}. A timing solution for PSR~J1740+1000,
 derived from 684 TOAs spanning
almost two years, is presented in Table 1. With a spin-down age $\tau_{s}$ of 114 kyr, this pulsar
is younger than 95\% of all pulsars listed in the Princeton Pulsar Catalog \cite{taylor933}.
Given the 23.85 pc cm$^{-3}$ DM of this pulsar, we use the Taylor \& Cordes (1993) model for
Galactic electron
density (hereafter TC93) to derive a distance of 1.4 $\pm$ 0.4 kpc.

Repeated observations
have shown that this pulsar's average flux density appears to be {\it higher}
at 1.4 GHz than at 430 MHz, with a spectral
index $\alpha$ (where $S_{\nu} \propto \nu^{\alpha}$ for an observing frequency $\nu$)
 of $0.9 \pm 0.1$.
We note that the spectral index estimate is 
highly contaminated by
interstellar scintillation, especially refractive
interstellar scintillation (RISS) which is uncorrelated between
the two frequencies.
 We note in passing that flat spectral indices may be characteristic
 of young pulsars \cite{lorimer95}, though the Crab is a counter example to this.

As shown in Figure 1, the 1.4 GHz profile is composed predominantly of
two conal pulse components,
while at 430 MHz a core component is apparent.
In Figure 2, we present  a polarization profile based on WAPP observations
carried out at 1475~MHz. The degree of linear polarization is 96\%.
Other pulsars with degrees of linear
polarization greater than 80\% include
J0134$-$2937, B0136$+$57, J0631$+$1036, B0656$+$14, J0742$-$2822, B0740$-$28,
B0833$-$45, B0906$-$49, B1259$-$63, J1359$-$6038, J1643$-$1224, B1737$-$30, B1913$+$167  and B1929$+$10
\cite{mcculloch78,wu93,man95,xilouris96,zepka96,gould98,man98,weisberg99}.
With some exceptions (most notably B1643$-$12 and B1913$+$167)
these pulsars have similar ages and/or periods to
PSR~J1740+1000, 
supporting a possible correlation between the degree of
linear polarization and period and/or age, as has been noted before
\cite{man71,mcculloch78,von98}. From data taken over a 100-MHz bandwidth at 1.475 GHz, we
derive a rotation measure of 23.8 $\pm$ 2.8 rad m$^{-2}$. For a DM of 23.85 pc cm$^{-3}$,
this translates to an average magnetic field along the line of sight of 1.2 $\pm$ 0.1 $\mu$Gauss.  
 
Because the pulsar's young age, estimated distance and high Galactic latitude 
suggest a high space velocity,
a proper motion measurement is very desirable. However, the
pulsar's large timing residuals
(see Figure 3), currently prohibit this.
The scatter in these residuals is likely  dominated by the considerable pulse-to-pulse jitter exhibited
by this pulsar, as shown in
Figure 4. At 1.4 GHz, 
both single-peaked and multiple-peaked pulse shapes
are recognizable, with the single-peaked modes generally stronger.
Sometimes a precursor, arriving $\sim$ 10~ms
before the main pulse, is present. More infrequently, a stronger postcursor,
arriving $\sim$ 10 ms after the main pulse, is apparent.
An auto-correlation analysis shows that pulses 
are intrinsically modulated on timescales $\sim$ 1 s (i.e. 6-7 pulses), as shown in 
Figure 5. This timescale
is similar at both 1.4 GHz and 430 MHz. 
The modulation index $m$,
corrected for additive off-pulse noise, 
\begin{equation}
m = \frac
	{(\sigma_{\rm on}^{2} - \sigma_{\rm off}^{2})^{1/2}}
	{<I_{\rm on}> - <I_{\rm off}>},
\end{equation}
where $\sigma_{\rm on}$ and $\sigma_{\rm off}$ are the on- and off-pulse RMS and
$<I_{\rm on}>$ and $<I_{\rm off}>$ are the average on- and off-pulse intensities, 
is calculated across the pulse and shown in Figure 6. 
The shape
is similar to that calculated for other
strongly modulated pulsars, with greater modulation on the outer edges of the pulse
\cite{backer73,bartel80}.
 
\subsection{Velocities}
 We define the $z$-velocity of an object of ``kinematic'' 
age $t$ born at height $z_0$ above the Galactic plane  as
\begin{equation}
V_{z} = \frac{z - z_{0}}{t},
\end{equation}
for a present-day height of $z= D\sin b$. For PSR~J1740+1000, 
assuming $z_{0}$ = 0, 
$D$ = 1.4 kpc and $t = \tau_s$,
we find $V_{z}$
= 4160 km s$^{-1}$.
Even for a $z_{0}$ of 0.3 kpc, over twice the best-fit scale
height of the pulsar progenitor population \cite{cher98,arz01},
a velocity of 1500 km s$^{-1}$ is
required for PSR~J1740+1000 to have reached its present-day $z$ of 0.48 kpc in its
114-kyr lifetime. 
Note that,  even for velocities as large as 4000 km s$^{-1}$, contamination of
$\tau_s$ due to the Shklovskii effect \cite{sh70} is insignificant, producing an apparent
$\dot{P}$
 of roughly
1\% of the true value.
 We later discuss the possibility that the kinematic age is much
larger than the spin-down age.
In Figure 7, we show $V_{z}$ (for $z_0$ of zero)
for all pulsars in the Princeton Pulsar Catalog \cite{taylor933} with latitude
 greater
than 10 degrees. We chose this latitude cutoff to exclude pulsars
within the progenitor
scale height and pulsars seen through much of the Galactic disk, where uncertainties in 
 TC93 are greater.

To relate an object's $z$-velocity to its transverse velocity
along of the line of sight, we adopt an
($x,y,z$) coordinate system such that $x,y$ define the Galactic plane, 
with $x$ towards longitude $l$ = 90 and $y$ towards $l$ = 180, and 
$z$ is toward the North Galactic pole. We then let 
$\hat{\bf n}$, $\hat{\bf l}$ and $\hat{\bf m}$ be unit vectors, 
with $\hat{\bf n}$ along the line
of sight to the pulsar, $\hat{\bf l}$ and $\hat{\bf m}$ 
transverse to the line of sight,
and $\hat{\bf l}\cdot \hat {\bf z} = 0$,
as in Eq. 28 of Cordes \& Rickett (1998) (hereafter CR98).
The pulsar's true space 
velocity in the $z$ direction is  
composed of radial and perpendicular components such that
\be
V_{z} = V_{r,z} + V_{\perp,z}
      = V_{r}\sin{b} + V_{\perp,m}\cos{b},
\ee
where $V_{r,z}$ and $V_{\perp,z}$ are the $z$ components of the pulsar's radial and transverse
velocity, respectively, 
and $V_{\perp,m}$ is the transverse velocity in the $\hat{\bf m}$ direction. 
Assuming no radial velocity and $z_0 = 0$, we find
$V_{\perp,m}$ =  4430 km s$^{-1}$ for PSR~J1740+1000. This is much 
higher than the
transverse velocity of
1700 km s$^{-1}$ estimated for the fastest known pulsar, PSR B2224+65 
\cite{cordes93}.
However, it is entirely possible that the space velocity is much
less than 4430 km s$^{-1}$, as we discuss below.  

\subsection{Interstellar Scintillation}

One way to get an independent constraint on a pulsar's velocity is through 
measurement of its
interstellar scintillations (ISS).
Because scattering by density inhomogeneities in the ionized interstellar
medium causes multiple rays to interfere, it produces frequency
structure in pulsar spectra
\cite{sch68,ric90}. These spectra change significantly on timescales of minutes to hours due to the 
high transverse
velocities of most pulsars.
When the spectrum is monitored over time, the resulting two-dimensional array or
``dynamic spectrum''  is often dominated by a random pattern of
scintillation maxima with characteristic bandwidth and timescale
\cite{ric90}. 
Intensity variations in time and frequency with scales
of a few minutes and a few MHz are due to diffractive ISS (DISS), 
caused by small-scale irregularities in the interstellar plasma. Longer
term intensity variations are caused by refractive ISS (RISS), 
due to variations in
electron density on much larger scales than those responsible for DISS. 

A two-dimensional correlation analysis can be used to calculate the
scintillation bandwidth $\Delta\nu_{d}$ and the scintillation timescale $\Delta t_{d}$,
which quantify the 
interstellar scintillation properties of a pulsar.
In Figure 8, we show an ACF of a dynamic spectrum for
PSR~J1740+1000. In Table 2 we list
diffractive bandwidths and timescales measured from ACFs
at various epochs. The average values of these parameters are
$\Delta t = 271 s$ and $\Delta \nu$ = 8.0 MHz, with standard deviations
of $\sigma_{\Delta t}$ = 146 s  and $\sigma_{\delta\nu}$  = 6.6 MHz .
The large variance in the scintillation parameters is likely due
to a combination of 
measurement uncertainties (from small numbers of scintillation maxima
in a given measurement) and
modulation from RISS, which may be enhanced by refraction from
the North Polar Spur (see below). 

We may use these measurements to
 estimate the speed of the ISS diffraction pattern with respect to the observer.
For a statistically uniform, Kolmogorov scattering medium, this speed
can be calculated as (CR98, Eq. 13)
\begin{equation}
V_{\rm ISS,5/3,u} = A_{\rm ISS,5/3,u} \frac{\sqrt{D\Delta\nu_{d}}}{\nu\Delta t_{d}}
\end{equation}
where $A_{\rm ISS,5/3,u}$ = $2.53\times10^4$ km s$^{-1}$ and $D$, $\Delta\nu_{d}$, $\nu$ and
$\Delta t_{d}$ have units of kpc, MHz, GHz and s, respectively. 
We list $V_{\rm ISS,5/3,u}$ for all epochs in Table 2 and, for a distance of 1.4 kpc,
 find an average value
of 216 km s$^{-1}$, with a standard deviation of 67 km s$^{-1}$.

For a uniform, Kolmogorov medium, $V_{\rm ISS,5/3,u}$ will be
equal to the pulsar's transverse speed.
For other cases, making the reasonable assumption that the speeds of
the observer and the medium are small compared to the pulsar's,
we may express the transverse speed of the pulsar as (CR98, Eq. 23)
\begin{equation}
V_{\perp} = W_{\rm C} W_{\rm D,PM} V_{\rm ISS,5/3,u},
\end{equation}
where
 the weighting factors $W_{\rm C}$ and $W_{\rm D,PM}$ relate the measured scintillation speed
to the pulsar's actual transverse
speed. The contribution of $W_{\rm C}$, which depends on both the wavenumber spectrum 
and the distribution of the medium (CR98, Eq. 18), is small (maximum range of 0.9-1.3).
We therefore
assume it is unity throughout the remaining analysis. The other factor,
$W_{\rm D,PM}$, is determined  by the distribution of the scattering material along the line
of sight to the pulsar.
This is proportional to $C_{n}^{2}$,
the coefficient of the electron-density wavenumber spectrum. 
An exact
expression for $W_{\rm D,PM}$ for a square law phase structure function is given in CR98 (Eq. 25) as
\be
W_{\rm D,PM}(D) &=& \left[\frac{2\int_{0}^{D} ds (s/D) (1 - s/D) C_{n}^{2}(s)}{\int_{0}^{D} ds (1 - s/D)
^{2} C_{n}^{2}(s)}\right]^{1/2}
\nonumber \\
&=& 	
	\frac
		{SM_{\tau}^{1/2}} 
		{
		\left [
		3\mbox{SM} - (\mbox{SM}_{\tau} + \mbox{SM}_{\theta})
		\right ]^{1/2} 
		}
\ee
where
$s$, the distance measured from the pulsar to the observer, varies from 0 to the pulsar
distance $D$ and
\be
\mbox{SM} &=& \int_0^D ds\, C_{n}^{2}(s) \\
\mbox{SM}_{\tau} &=& 6 \int_0^D ds\,(s/D)(1-s/D) C_{n}^{2}(s) \\
\mbox{SM}_{\theta} &=& 3 \int_0^D ds\, (s/D)^2 C_{n}^{2}(s).
\ee
define the scattering measure, the scattering measure weighted for pulse broadening,
and the scattering measure weighted for angular broadening, respectively.

If we use TC93 to calculate SM, SM$_{\tau}$ and SM$_{\theta}$ and thus the weight factor
$W_{\rm D, PM}$, we find $W_{\rm D, PM}$ = 0.85 and $V_{\perp}$ = 184 km s$^{-1}$, far
smaller than the kinematic $z$-velocity calculated using
$z_0 = 0$ and $t = \tau_s$.  
In the following, we consider several ways to reconcile
the various constraints on the pulsar's velocity.  Some 
reduce the value of the kinematic $z$-velocity while
others increase the scintillation estimate of the transverse
velocity. We note that, if the actual distance to the pulsar is much less than the
distance estimated by TC93, the difference between the velocity estimated
from ISS (proportional to $\sqrt{D}$) and the kinematic velocity (proportional to $D$)
 becomes smaller.

\section{Interpretation of J1740+1000's Kinematics}

\subsection{Large Radial Velocity}

An unrealistically large radial velocity is required if we assume
that it accounts solely 
for the difference between the $V_z$ calculated
assuming $V_r = 0$ and the ISS speed, which reflects only the transverse
components. Letting $z$ be the pulsar's height above the Galactic plane, $z_{\rm TC}$ the height above
the plane given
the TC93 model distance and ignoring any contribution to
$V_z$ from $\vperp$, we have
\be
V_r &\approx& 
	\left( \frac{D_{\rm TC}}{\tau_s} \right )
	\left( \frac{\tau_s}{t} \right )
	\left( \frac{z-z_0}{z_{\rm TC}} \right)
\nonumber \\
    &\approx&
	1.2 \times 10^4 \,{\rm km \,\,s^{-1}}
	\left( \frac{\tau_s}{t} \right )
        \left( \frac{z-z_0}{z_{\rm TC}} \right).
\ee
Assuming that a value $V_r\approx 10^3$ km s$^{-1}$ is ``reasonable''
in comparison with the highest known velocities, it appears
that either the pulsar is  kinematically much older
than the spindown age such that $t\gg\tau_s$,  or the pulsar was born
at a relatively high $z_0$ such that $\vert z - z_0 \vert \ll z_{\rm TC}$.
However, we later discuss other interpretations where the transverse velocity makes
a sizeable contribution to $V_z$ and $V_r$ need not be so large.

\subsection{Large Kinematic Age}

Depending on the spindown law for the pulsar
and on the evolutionary history of the neutron star's progenitor, 
the spindown age $\tau_s$ may or may not be a good estimate for the kinematic
age $t$.
First, the spindown age may only approximate the true age of the  pulsar.
Young pulsars show estimated braking indices
$n \equiv \Omega\ddot\Omega / \dot\Omega^2 < 3$, where $\Omega = 2\pi/P$, implying ages greater 
than the spindown age $\tau_s$, which is  calculated for $n=3$.  
However, a pulsar born with spin period not significantly smaller 
than its present-day period could be much younger than $\tau_s$.   
In fact, a statistical comparison of kinematic $z$-velocities and proper motion
derived velocities suggests that, on average, pulsars are about
50\% {\it younger} than their spindown ages (Cordes \& Chernoff 1998).

Recently, Gaensler \& Frail (2000) argued that the pulsar
PSR~B1757$-$24 was significantly {\it older} than $\tau_s$ by comparing
the pulsar's location to its birth place, estimated through identification
of the geometric center of the associated supernova remnant.   
The upper bound on the proper motion of the synchrotron nebula around the
pulsar suggests an older age. However,  we believe that the errors associated
with estimating ages based on remnant morphology are too large to allow
any general statement about pulsar ages.

Secondly, even if the pulsar age is well approximated by $\tau_s$, the kinematic
age could be significantly larger.  For example, the progenitor
of J1740+1000 may have been a member of a binary system whose primary
exploded long before the observed pulsar was produced, imparting
significant space velocities to its remnant star and to the progenitor
of J1740+1000, either as an intact binary or as individual
objects.   A scenario of this type was proposed by 
Gott, Gunn \& Ostriker (1970)
to account for the current locations of the
Crab pulsar (B0531+21) and the nearby, older and longer-period pulsar,
B0525+21, with the binary producing both pulsars
originating in the I Gem OB association.     

\subsection{Halo Progenitor}

It is conceivable that 
PSR~J1740+1000 may have been born well out of the Galactic plane
from a halo-star progenitor, perhaps through a rare accretion induced
collapse (AIC) or merger event \cite{bail90,ras95}.  If the neutron star formed through
 accretion induced collapse (AIC) of a halo white dwarf,
the expected transverse speed would be $\sim 250$ km s$^{-1}$
(i.e. the characteristic speed of a halo star),
in the absence of a supernova-induced kick. This is consistent
with the scintillation estimate.
Given one object in 114 kyr, a na\"{\i}ve estimate of the birth rate
of such objects would be $\sim 10^{-5}$ yr$^{-1}$.  This rate
is about $10^3$ times smaller than that of
disk-born pulsars and about
$3-10$  times larger than the estimated birth rate of
millisecond pulsars (Lorimer et al. 1995; Cordes \& Chernoff 1997). 
Therefore, if  AIC produces only large magnetic
field objects, the
birth rate implied by PSR~J1740+1000 would be compatible with
birth rates of other neutron star subpopulations. Because we have
not accounted for any selection effects in detecting objects produced through AIC,
this is of course a very rough estimate.

\subsection{Interstellar Medium Structure and DISS}

The local ISM has been sculpted by supernovae and stellar winds into
shells and loops that could perturb both the pulsar's distance estimate,
which is based on the dispersion measure, and the scintillation
parameters.    If only the distance is in error, reconciliation of
the DISS and kinematic velocities implies $D\approx 4$ pc (for $z_0 = 0$), 
which requires an unrealistically  large HII column density 
within a few parsec of the Sun.   For $z_0\ne 0$ the distance could be
larger and a smaller fraction of DM would be needed from an HII region.   

As shown in Figure 9,
the line of sight to PSR~J1740+1000
passes through the North Polar Spur (NPS) and the Gould Belt,
an expanding disk of gas and young stars. The North Polar Spur is
the brightest part of the larger radio feature called Loop I, a
116$^{\circ}$ diameter circular feature on the sky centered at longitude
 330$^{\circ}$ and 
latitude +18$^{\circ}$. 
The NPS rises from the Galactic plane at a
longitude of $30^{\circ}$
and covers latitudes of $15-30^{\circ}$, appearing as a bright, narrow
ridge in radio and X-ray maps.  Sofue (1977)
postulates that the NPS is part of a shock front induced by an explosive event at the Galactic center.
However, 
a local, supernova remnant origin for the NPS is usually assumed 
\cite{heiles80,salter83,egger95,heiles98}. 
Distance estimates given this assumption range from 50 to 200 pc \cite{berk73,bingham67}, while the thickness
of the NPS is estimated to be $\sim$ 20 pc \cite{berk73}. 
The distance of the Gould Belt in the direction of PSR~J1740+1000
is 100 to 300 pc, also much less than the TC93  
distance for PSR~J1740+1000.

The discovery of PSR~J1740+1000 may have implications for the 
long-standing question of how the NPS
was formed. While it is generally accepted that such shells 
are produced through the release of energy into the interstellar
medium, the age of the NPS is uncertain. Its slow rate of 
expansion
  indicates that it is fairly old, but 
the production of X-rays in the interior poses an apparent contradiction. If these X-rays
are due to reheating processes, the NPS could be roughly $2\times10^{6}$ years old
and produced by one normal supernova explosion. However, if the
X-ray emission is due to just a single event, the age of the NPS could be only $2\times10^{5}$ 
years \cite{heiles80}, roughly the spin-down age of PSR~J1740+1000.

Structure in the local ISM like that seen in the NPS and Gould Belt
perturbs both
DM and SM for the line of sight to the pulsar.  Strong evidence for the diffracting
and refracting power of the NPS comes from examining the distribution of Extreme Scattering
Event (ESE) sources,
extragalactic sources which show extreme variations in their radio light curves.
ESEs are generally believed to be caused by strong diffraction and refraction
due to enhanced electron density turbulence in the ISM.
As shown in Figure 9, three of the ten known ESE sources (1749+096, 1821+107 \& 1741$-$038)
have lines of sight which pass near or through the North Polar Spur \cite{fey96}.

To proceed, we follow
Chatterjee et al. (2001) by superposing a thin screen at distance $D_s$ from
the pulsar on a smoothly varying
medium such as is modeled by TC93.   This yields a
DISS velocity correction factor 
\be
W_{\rm D,PM}(D) &=& 
\frac
	{
	\left[
		6(D_s/D)(1-D_s/D) \Delta\mbox{SM} + SM_{\tau}
	\right]^{1/2}
	}
	{
	\left[
	3 (1-D_s/D)^2 \Delta\mbox{SM}
	+ 3\mbox{SM} - (\mbox{SM}_{\tau} + \mbox{SM}_{\theta})
	\right]^{1/2}},
\ee
where $\Delta$SM is the additional scattering measure contributed 
by the screen. 
Given the standard definitions of SM and DM
(i.e. Eq. 9 and DM = $\int_{0}^{D} n_{e}(s) ds$, where $n_{e}$ is the electron
density in the medium), 
we may relate $\Delta$SM and $\Delta$DM,
 the differential
DM contributed by the screen. For a screen of width $\Delta$s, 
\begin{equation}
\Delta\mbox{SM} = (1/3)(2\pi)^{-1/3} K_{u} F_{s}
\frac{(\Delta\mbox{DM})^2}{\Delta s},
\end{equation}
where $K_{u} = 10.2\times10^{-3}$m$^{-20/3}$cm$^{6}$ kpc pc$^{-1}$ is a unit
conversion factor, 
and the units of $\Delta\mbox{SM}$,
$\Delta\mbox{DM}$, and $\Delta s$ are kpc m$^{-20/3}$, pc cm$^{-3}$ and pc, respectively.
As defined in TC93, $F_{s}$ is a dimensionless
``fluctuation parameter'' (i.e. $C_{n}^{2}(s) \propto F_{s} n_{e}^{2}(s)$) 
which is found to  range from 0.1 to 100
for the various Galactic components of the TC93 model. 
We note that while DM and SM are correlated, 
the excess electron density from a screen could cause a
large fractional change in scattering measure but a much
smaller fractional change in dispersion measure.

Given the SM, SM$_{\tau}$, and SM$_{\theta}$ estimated by TC93 along the line of
sight to the pulsar, $W_{D,PM}$ will increase with increasing $\Delta$SM,
asymptoting to a value of $[2(D_{s}/D)/(1-D_{s}/D)]^{1/2}$. For the measured ISS
velocity to be made consistent with $z_{0} = 0$ (assuming $t = \tau_{s}$ and ignoring the
radial velocity contribution to $V_{z}$), a
$W_{D,PM}$ of 20.5 is needed. This requires an unreasonable
$D_{s}/D = 0.995$, or, for a pulsar distance of 1.4 kpc, a
screen that is only 7 pc away from the Earth. To calculate the minimum $z_{0}$ for a
range of realistic model parameters,
we constrain $\Delta$SM such that the resultant $\Delta$DM is
less than half of the pulsar's DM. We note that 
 $\Delta \mbox{DM} =  11.9$ pc cm$^{-3}$
corresponds to a screen electron density of 0.6 cm$^{-3}$ for a screen thickness of 20 pc,
roughly consistent with the constraint $n_{e}$ $\le$ 0.4 cm$^{-3}$
derived by Heiles et al. (1980) 
from NPS rotation and emission measures.
Assuming $\Delta$s, $D_{s}$, $D$, and $F$ of 20 pc, 130 pc, 1.4 kpc, and 10,
we calculate a minimum $z_{0}$ of 0.34 kpc.
Allowing screen distances as small as 50 pc, a pulsar
distance as low as 0.6 kpc, a screen thickness as small as 10 pc,
and a fluctuation parameter $F_{s}$ 
as large as 1000, 
the lowest height at which the pulsar could have been
born is 0.10 kpc, consistent with estimated distances to the NPS. Note that these minimum
values of $z_0$ depend on the assumption that there is no radial
contribution to the transverse
velocity and that 
the kinematic age is equal to the spin-down age. 

\subsection{Refractive Modulation of DISS}

Another effect of a strong phase screen is that it can refract
the radiation into a cone of received directions that is not centered
on the direct ray path between pulsar and observer.  The result
is to decrease the DISS bandwidth while leaving the DISS time scale
unchanged (Cordes, Pidwerbetsky \& Lovelace 1986). 
It is therefore possible that strong RISS is
contaminating our measured $\Delta\nu_{d}$, causing an underestimate of
$V_{ISS,5/3,u}$. For the $V_{ISS,5/3,u}$ estimate to agree with the $V_{\perp}$
estimated assuming $V_{r} = 0$, $z_{0} = 0$ and $t = \tau_{s}$, the actual
bandwidth, unperturbed by RISS, would have to be an impossibly high 3360 MHz.
However, while RISS may not be solely responsible for the discrepancy between 
the pulsar's age and height above the plane and the measured scintillation parameters,
it could be biasing our results.
The distribution of ESE sources provides strong evidence for the refracting properties of
the NPS.

\section{Prospects for Astrometry}

The most robust way to determine the velocity and possible origin of 
PSR~J1740+1000 is through measurement of a proper motion and parallax. 
 An estimated distance of 1.4 kpc corresponds
to a parallax of 0.71 mas. A transverse speed of 
$10^3$ km s$^{-1}$ for a pulsar at that distance corresponds
to a  proper motion  $\sim 150$ mas yr$^{-1}$.

Because the timing residuals of the pulsar are large, 
the prospects for measuring the proper motion through VLBI are much better than through
timing.
Given the current RMS timing residuals, we estimate that
we could expect to measure a 150 mas yr$^{-1}$ proper motion
to about 10 sigma with two more years of timing data. 
However, if there is significant
timing noise, as is expected for a young pulsar, 
a proper motion may be unmeasurable.
A parallax measurement through timing is not feasible.

The current Very Large Baseline Array (VLBA) has sufficient angular
resolution to easily measure the anticipated proper motion. 
However, the $\sim 2$ mJy 1.4-GHz average flux density  of this pulsar
necessitates more collecting area than is available with the VLBA
alone.  With the addition
of Arecibo and the GBT to the array, a proper motion should be 
measurable within several months.
Moreover, we have seen episodes of scintillation enhancement of the
pulsar's flux density for sustained times ($\sim 1$ hr) by a factor
of ten, which will boost the astrometric sensitivity of VLBI 
measurements.
As in the case of PSR~B0919+06 \cite{shami01}, the situation may also be  aided
by the presence of two nearby calibrators with separations and flux
densities of
4 and 5 arcminutes and 2.7 and 2.6 mJy, respectively \cite{condon98}. 
The in-beam calibrators may also allow measurement of the pulsar's
parallax, especially if the pulsar is associated with the nearby NPS or Gould Belt.

\section{High-Energy Emission}

Because of its youth and proximity, PSR~J1740+1000
is an excellent candidate for detection at high energies. Assuming that
X-ray flux is proportional to $\dot{E}/D^{2}$ \cite{becker97}, where $\dot{E}$ is
the spin-down energy loss rate,
PSR~J1740+1000 has the 10th highest predicted X-ray flux out of 547 pulsars in the Princeton Pulsar Catalog
\cite{taylor933}.
 Upcoming observations with \it Chandra \rm may reveal the
existence of thermal or magnetospheric radiation from the star. Nebular emission from a
compact synchrotron nebula or a pulsar wind nebula bow-shock, especially likely if the
velocity of the pulsar is high,
should be detectable.

This pulsar should also be a strong gamma-ray source.
Given the model of McLaughlin \& Cordes (2000), PSR~J1740+1000
ranks 20th out of 547 pulsars for OSSE (50$\le E \le$200 keV) flux and 10th out of 547 pulsars
for EGRET ($E>$100 MeV) flux.
 Because this pulsar was not detected as an EGRET point
source \cite{hart99},
we can place an upper limit on its gamma-ray flux of $8.7\times10^{32}$ ergs s$^{-1}$ kpc$^{-2}$,
just above the flux of $8.1\times10^{32}$ ergs s$^{-1}$ kpc$^{-2}$ predicted by the
McLaughlin \& Cordes model. Unfortunately, even in the closest EGRET viewing period, 
PSR~J1740+1000 is 20 degrees off-axis. Still, we did search this and other viewing periods
for periodicities over a range of periods and period derivatives close
to that predicted by our current timing model. To do this, we first
created a barycentered
time series of photon arrival times given the known position of the pulsar. Photons
are then individually weighted by the EGRET point spread function and folded for each trial
ephemeris. The point spread function weighting procedure produces a dramatic improvement in
signal-to-noise. For each profile, we use the Bayesian procedure presented in McLaughlin et al. (1999)
to calculate the probability of a pulse of unknown width and phase. Using this method, we detect the known
EGRET pulsars with high signal-to-noise. While, not surprisingly, we find no convincing pulsed
signatures for PSR~J1740+1000,
this pulsar should be easily detectable, modulo any beaming effects,
with the Gamma-Ray Large Area Space Telescope (GLAST), with a projected sensitivity of
$3\times10^{31}$ 
ergs s$^{-1}$ kpc$^{-2}$.
 Gamma-ray observations of this pulsar will be aided greatly by its high latitude,
where the 
Galactic gamma-ray background is much smaller than in the Galactic plane.

\section{Conclusions}
 
PSR~J1740+1000 is a young pulsar whose study will
yield important information about pulsar velocities
and the local interstellar medium.
As a young object at a high Galactic
latitude, its existence may signify a complex progenitor history, either
from a Galactic disk population or from a halo population. 
Alternatively, it may simply have an especially high velocity. 

Using simple models, it is difficult to reconcile the measured scintillation properties
of this pulsar with its age and height above the Galactic plane.
 While it is of course possible that this
pulsar was indeed born out of the Galactic plane, there are several ways in which
the measured ISS parameters could be reconciled with a birth in the midplane of the Galaxy.
Clearly the best way to
resolve this issue is through measurement of the  proper motion and 
parallax. With the upcoming addition of
Arecibo and the GBT to the VLBA, a proper motion
will be measurable within a few months. Measurement of the parallax will
rely heavily on the quality of nearby calibrator sources and exploitation
of strong scintillation maxima that boost the apparent flux density of the
pulsar by a factor of ten or more.  
We are also
planning high-energy observations with
{\it Chandra} and
GLAST, as, due to its youth and proximity, the pulsar should 
be a strong high-energy source.

\acknowledgments

We thank Shami Chatterjee for useful discussions.
This work was partially supported by NSF grant AST-9819931.
The work was also supported by the National Astronomy and Ionosphere Center,
which is
operated by Cornell University under cooperative agreement with the National
Science
Foundation (NSF). Part of this work was performed while ZA held a
        National Research Council Research Associateship Award at
        NASA-GSFC.  We  acknowledge 
 the use of NASA's SkyView facility
            (http://skyview.gsfc.nasa.gov) located at NASA Goddard
            Space Flight Center.

{}

\begin{deluxetable}{lccccccc}
\tablewidth{4.0in}
\tablenum{2}
\tablecaption{DISS Parameters Measured at 1.4 GHz for PSR~J1740+1000}
\tablehead{
\colhead{MJD} & \colhead{$\Delta\nu_{d}$ (MHz)} & \colhead{$\Delta t_{d}$ (s)} & \colhead{$V_{\rm ISS,5/3,u
}$ (\rm km \hspace{0.01in}  s$^{-1}$)}}
\startdata
51314& 1.8 & 85 & 329 \\ 
51674& 2.4 & 226 & 147 \\ 
51722& 2.0 & 193 & 157 \\ 
51851& 16.0 & 372 & 226 \\ 
51968& 5.4 & 211 & 231 \\ 
51986& 16.4 & 537 & 157 \\ 
52022& 11.7 & 273 & 262 \\ 
\enddata
\end{deluxetable}{}

\begin{deluxetable}{lr}
\tablewidth{0pt}
\tablecaption{Observed and Derived Parameters for PSR~J1740+1000}
\tablehead{Parameter & Value}
\startdata
Right Ascension (J2000)            & $\rm 17^{\rm h} 40^{\rm m}25\fs 950(5)$ \\
Declination     (J2000)            & $\rm +10^\circ  00'      06\farcs 3(2)$\\
Galactic longitude                 & $34.01067(7)^{\circ}$ \\
Galactic latitude                  & $+20.26828(5)^{\circ}$ \\
Barycentric Period (s)             & 0.154087174313(2) \\
Period derivative ($10^{-15}$ s s$^-1$)     & 21.4651(2) \\
Epoch (MJD)                        & 51662 \\
Dispersion Measure (pc cm$^{-3}$)  & 23.85(5)\\
Rotation Measure (rad m$^{-2}$)    & 23.8(2.8) \\
Timing data span (MJD)             & 51310--52014\\
Number of TOAs                     & 684 \\
Post-fit RMS timing residual ($\mu$s)     & 952 \\
\hline
Flux density at 0.4 GHz (mJy)      & 3.1(2) \\
Flux density at 1.4 GHz (mJy)      & 9.2(4) \\
Mean spectral index                & 0.9(1) \\
\hline
Distance\tablenotemark{\star}\, (kpc)                     & 1.4 \\
Dipole magnetic field strength\tablenotemark{\dagger}\, $B$ ($10^{12}$ G)&1.8\\
Characteristic age\tablenotemark{\dagger}\, $\tau_{s}$ (kyr)           & 114\\
\enddata
\label{tab:1740}
\tablecomments{Figures in parentheses represent $1\,\sigma$
uncertainties in least-significant digits quoted. }
\tablenotetext{\star}{Calculated using the TC93
model}
\tablenotetext{\dagger}{Calculated using the standard
magnetic dipole formulae viz: $B=3.2\times10^{19} \sqrt{P\dot{P}}$ Gauss;
$\tau=P/2\dot{P}$ (Manchester \& Taylor 1977)}
\end{deluxetable}

\clearpage

\begin{figure}[ht]
\epsscale{0.8}
\plotone{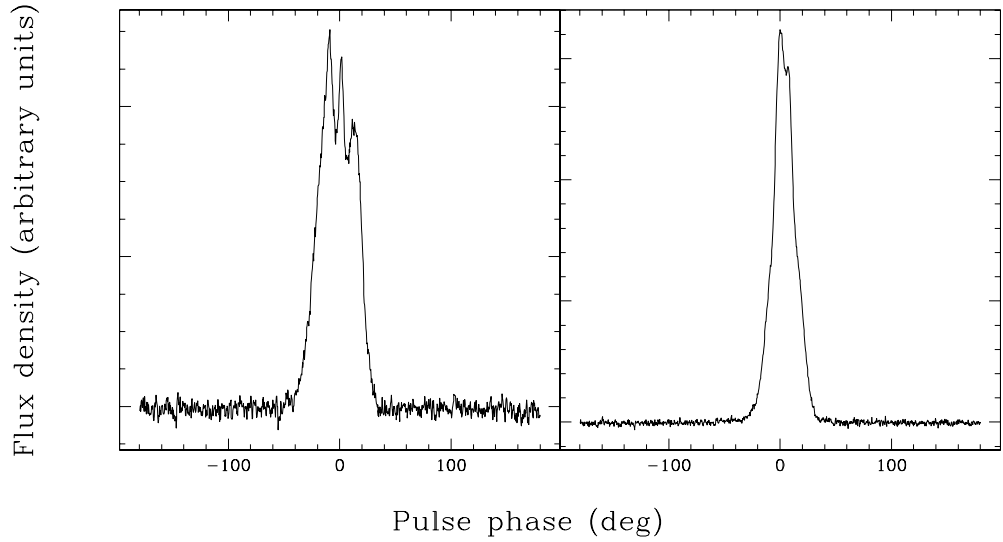}
\caption{Folded pulse profiles for PSR~J1740+1000 at 430 MHz (left) and 1.4 GHz (right).
 These profiles were formed by co-adding 17000 s and 11000 s of PSPM data at 430 MHz
and 1.4 GHz, respectively.
For all observations, data were taken across an 8-MHz
 bandwidth and
dedispersed offline using a DM of 23.85 pc cm$^{-3}$.}
\label{fig:profiles}
\end{figure}

\begin{figure}[ht]
\epsscale{0.8}
\plotone{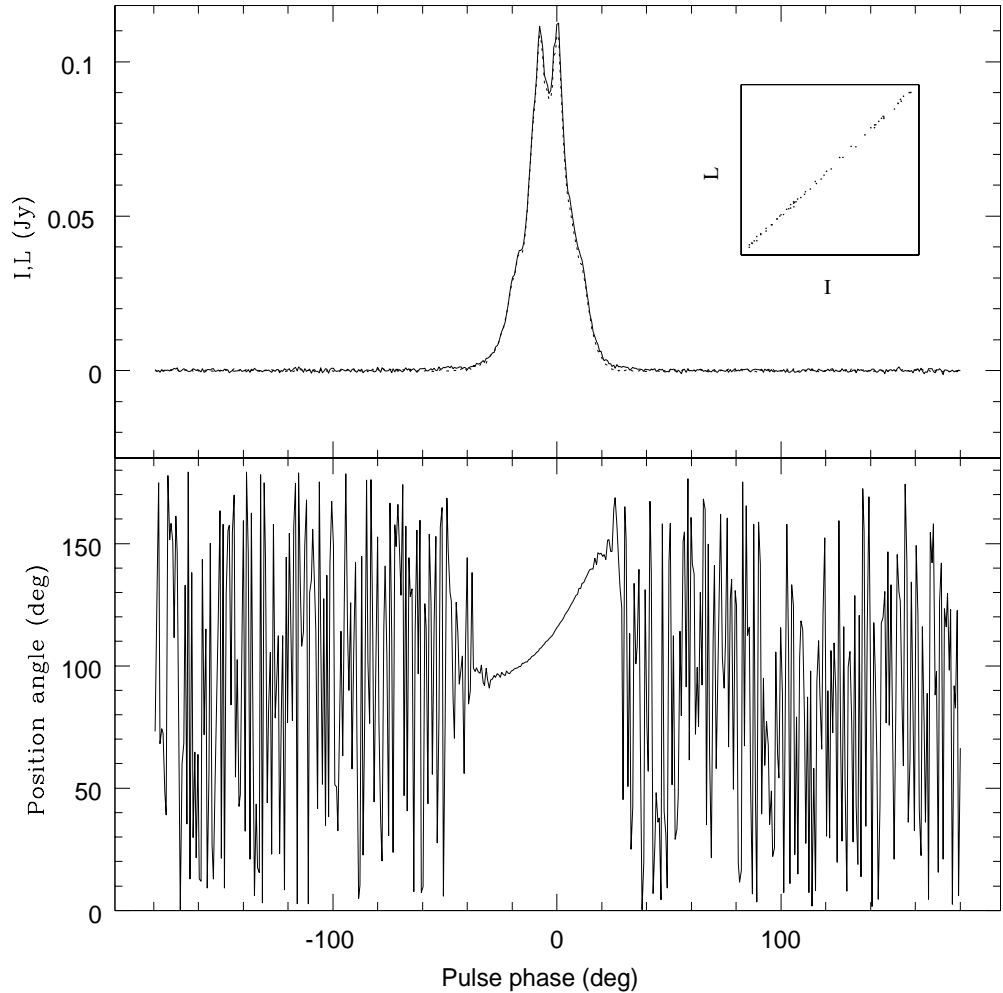}
\caption{Polarization profile of PSR~J1740+1000 at 1475~MHz. The upper panel shows
the total intensity I (solid line) and linearly polarized intensity L (dashed line). These
lines are nearly indistinguishable due to the high degree of linear polarization.
Because this profile was formed from only 4 minutes of data, the
pulse shape has not yet converged to that shown in Figure 1.The inset plot shows L vs. I across the pulse. The
lower panel shows the polarization
position
angle. We do not show the circularly polarized power because it is influenced
by the unknown cross coupling of the feed system.}
\label{fig:polprof}
\end{figure}

\begin{figure}[ht]
\epsscale{0.8}
\plotone{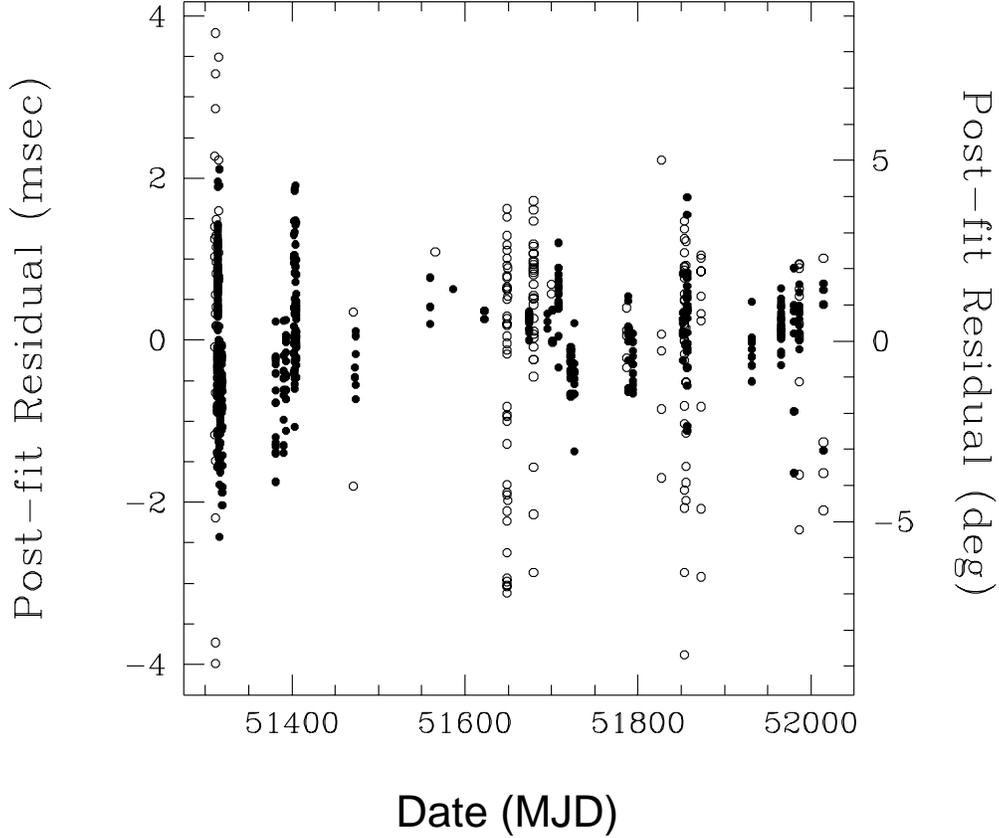}
\caption{Timing residuals for 684 TOAs spanning MJDs 51310 to 52014
for the timing model in Table 1, which includes a fit for
$P, \dot P$, DM, RA and DEC.  Fitting for proper
motion or a frequency second derivation improves the fit somewhat, but
since we are unsure of the origin of the long-term trend, we do not include
it here.
Open and closed circles denote measurements at 430~MHz and
1.4~GHz, respectively. TOA uncertainties range from 0.5 to 5 ms at 430~MHz
and from 0.3 to 3 ms at 1.4~GHz, varying by an order of magnitude or more
due to scintillation and profile shape variations.
 Note that, despite the variation in TOA uncertainity,
all TOAs were given equal weight in the least-squares fit to minimize biasing
from profile shape variations between individual scans.}
\end{figure}

\begin{figure}[ht]
\epsscale{0.5}
\plotone{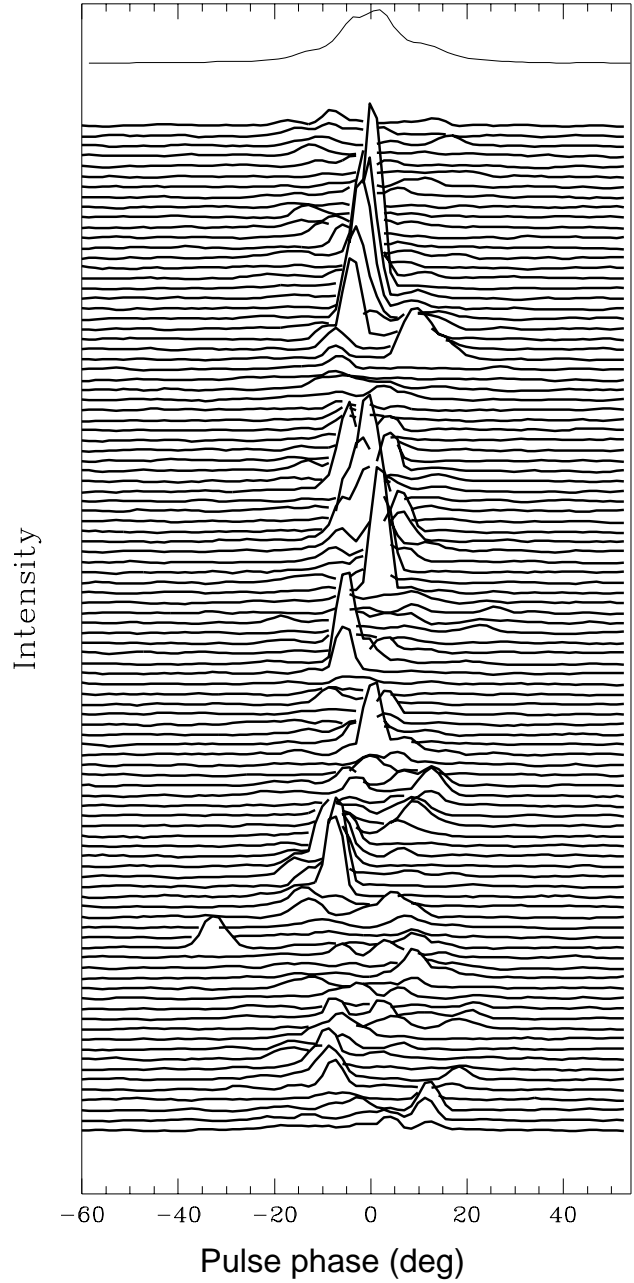}
\caption{A sequence of 100 single pulses (i.e. $\sim$ 15 seconds of data)
from PSR~J1740+1000. These data were taken using the WAPP
at 1.475 GHz across a 100-MHz bandwidth. The time resolution
is 0.8~ms.}
\end{figure}

\begin{figure}[ht]
\epsscale{0.8}
\plotone{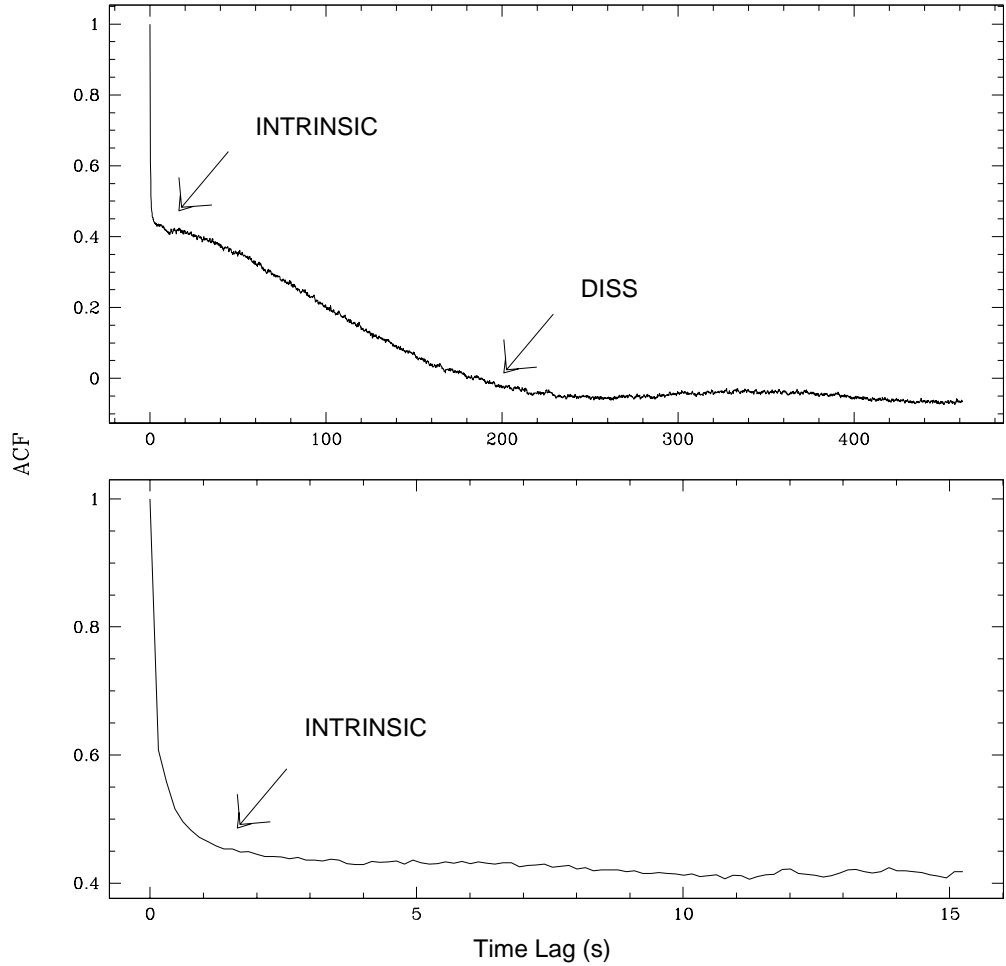}
\caption{Auto-correlation function (ACF) for a dedispersed time
series of period-averaged intensities for PSR~J1740+1000.
These data were taken with the
AOFTM at 1.4 GHz across a
10-MHz bandwidth.
The lower plot shows only the first 16 seconds.
The variation with timescale $\sim$ 200 s is caused by diffractive interstellar scintillation (DISS),
while the $\sim$ 1 s variation is due to intrinsic pulse-to-pulse
fluctuations.}
\end{figure}

\begin{figure}[ht]
\epsscale{0.8}
\plotone{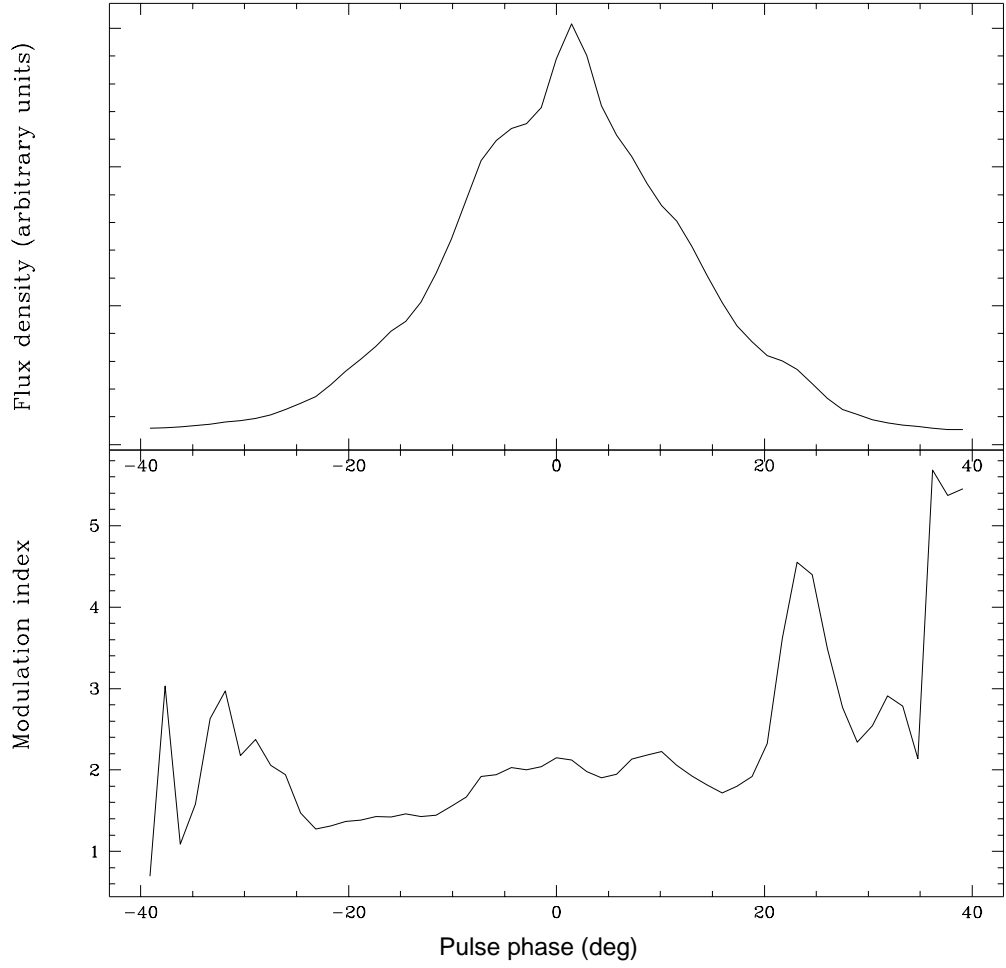}
\caption{The inner 20\% of the 1.475~GHz pulse profile, along with the corresponding
 modulation index, is shown for the full 1200-second observation from which the sequence
in Figure 4 was taken.
}
\end{figure}

\begin{figure}[ht]
\epsscale{0.8}
\plotone{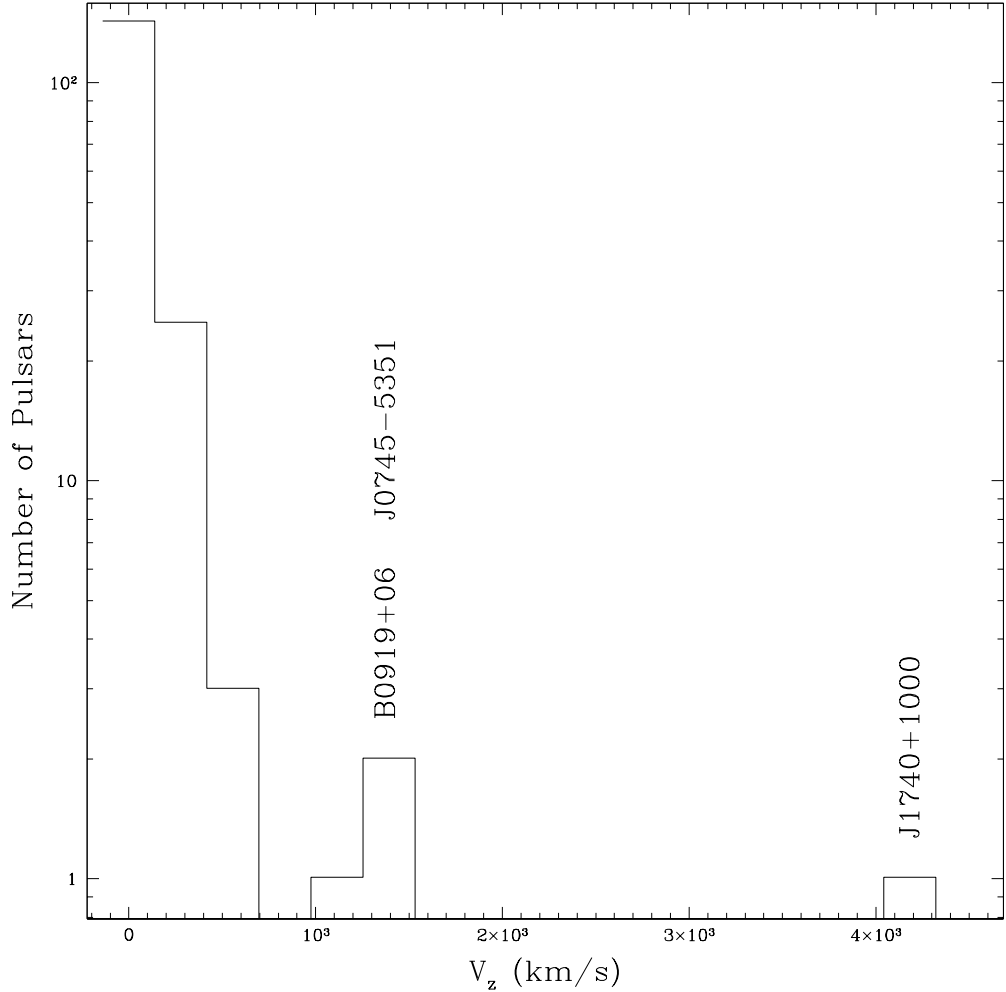}
\caption{$V_{z}$ is plotted for all pulsars
in the Princeton Pulsar Catalog \cite{taylor933}
with Galactic latitudes $\vert b \vert > 10^{\circ}$.
Note the logarithmic scaling of the y-axis.
The value for
PSR~J1740+1000 is anomalously high. Pulsar
distances were calculated using TC93, except in cases of  a more
accurate parallax measurement. We note that $V_{z}$ can be quite different
from measured transverse velocities. For instance,
PSR~B0919+06 has a $V_{z}$ of 1400 km s$^{-1}$, but an astrometrically-determined
transverse velocity of 505$\pm$80 km s$^{-1}$
\cite{shami01}.
We also note that imposing a latitude cutoff omits pulsars such as
PSR~B0531+21 and PSR~B1509-58, which have $V_{z}$s of
$1.6 \times 10^{5}$ and $5.6 \times 10^{4}$ km s$^{-1}$,
respectively, but, with known supernova remnant associations, were
certainly born at
$|z_{0}| > 0$.
}
\end{figure}

\begin{figure}[ht]
\epsscale{0.8}
\plotone{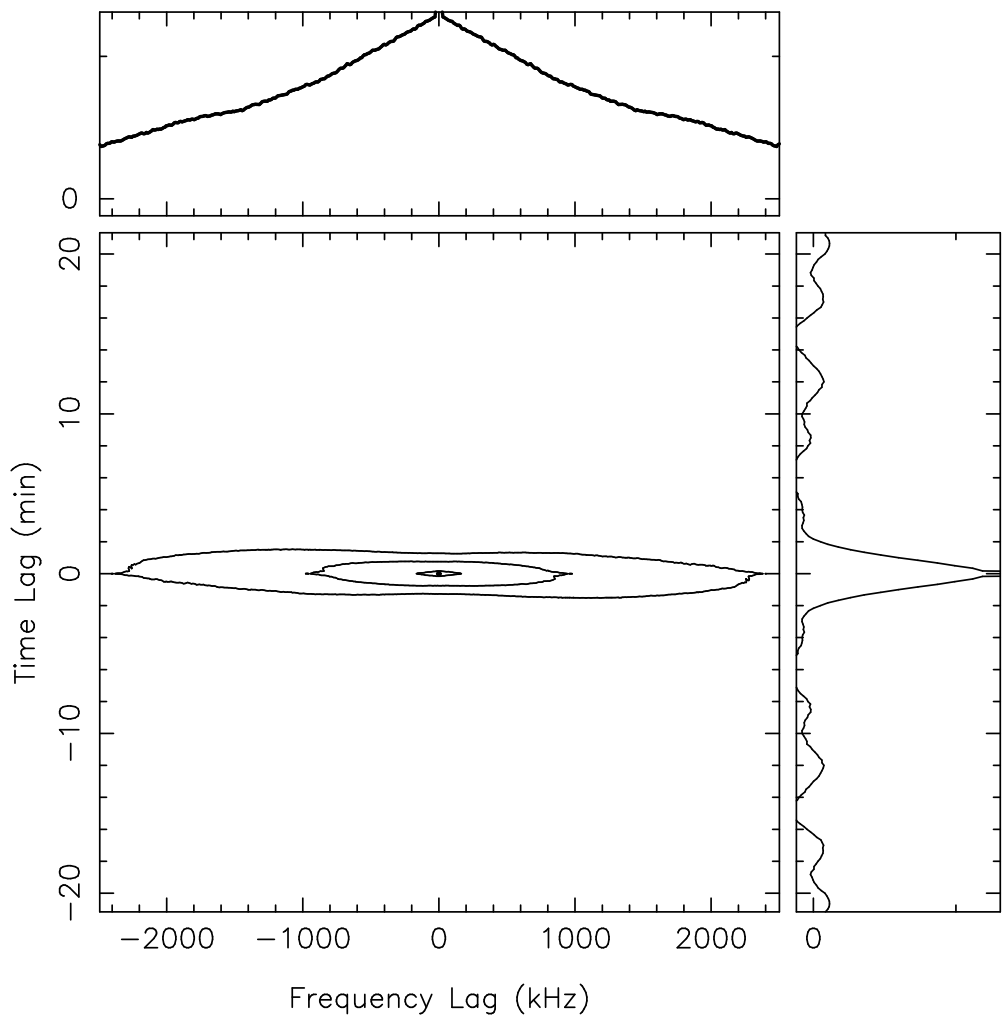}
\caption{A two-dimensional ACF for a dynamic spectrum for PSR~J1740+1000 taken at a center
frequency of 1410~MHz in May 1999. We also plot one dimensional slices along the two axes, used to
calculate $\Delta \nu_{d}$ (as the half-width
at half-max of the cut at zero time lag)
and $\Delta t_{d}$ (as the half width at $e^{-1}$ of the cut at
zero
frequency lag). For this epoch, we calculate
$\Delta \nu_{d}$ = 85 s and $\Delta t_{d}$ = 1.8 MHz.}
\end{figure}

\begin{figure}[ht]
\epsscale{0.8}
\plotone{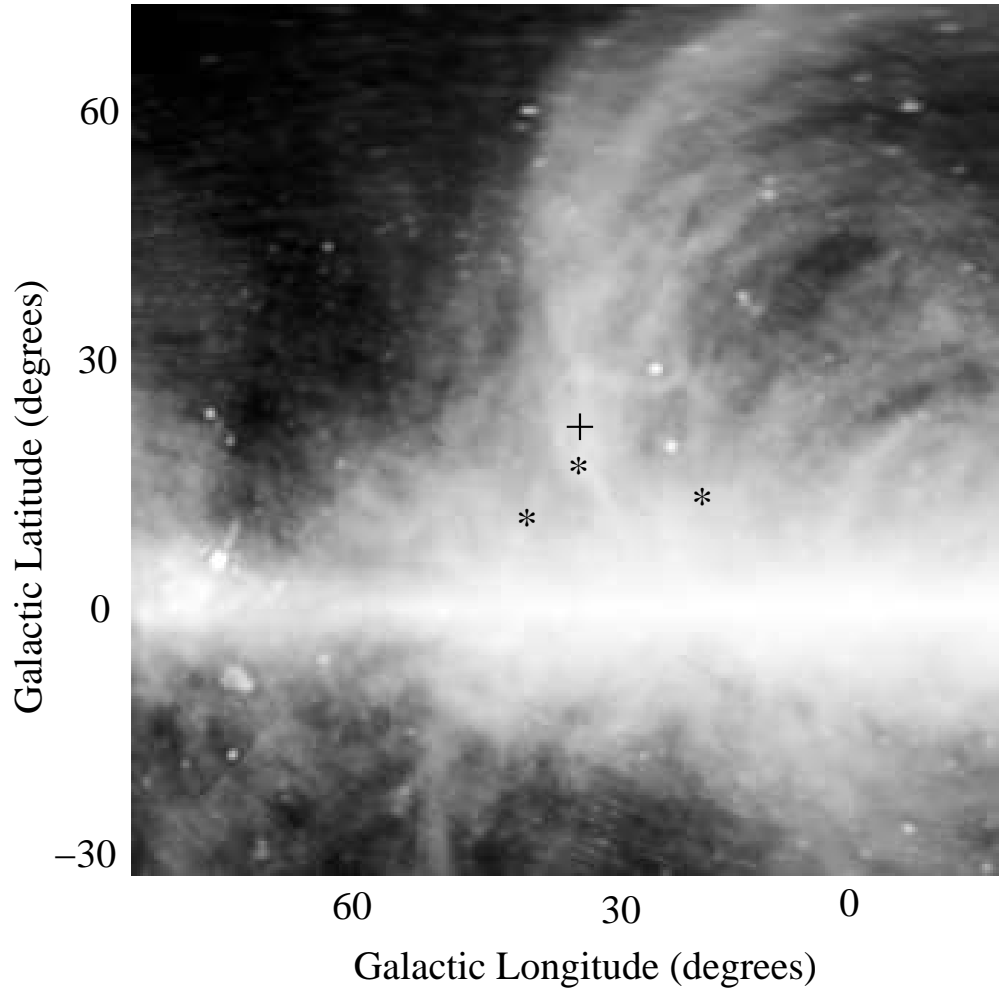}
\caption{A 105$^{\circ}\times105^{\circ}$ image of the Galaxy at 408 MHz,
showing the NPS rising from the Galactic plane  \cite{haslam82}.
The position of PSR~J1740+1000 is denoted with a cross. The positions of the
three Extreme Scattering Event Sources associated with the NPS are denoted with
asterisks.
}
\end{figure}

\begin{thebibliography}{}
\bibitem[Arzoumanian et al. 2001]{arz01} Arzoumanian, Z., Chernoff, D.\ F., \& Cordes, J.\ M., 2001, ApJ, submitted, astro-ph/0106159
\bibitem[Backer 1973]{backer73} Backer, D.\ C.\ 1973, \apj, 
182, 245
\bibitem[Bailyn \& Grindlay 1990]{bail90} Bailyn, C.\ D.\ \& 
Grindlay, J.\ E.\ 1990, \apj, 353, 159 
\bibitem[Bartel et al. 1980]{bartel80} Bartel, 
N., Sieber, W., \& Wolszczan, A.\ 1980, \aap, 90, 58  
\bibitem[Becker \& Tr\"umper 1997]{becker97} Becker, W.\ \& 
Tr\"umper, J.\ 1997, \aap, 326, 682
\bibitem[Berkhuijsen 1973]{berk73} Berkhuijsen, E.\ M.\ 1973, 
\aap, 24, 143
\bibitem[Bingham 1967]{bingham67} Bingham, R.\ G.\ 1967, \mnras, 
137, 157 
\bibitem[Chatterjee et al. 2001]{shami01} Chatterjee, S., Cordes, J.\ M.\, Lazio,
T.\ J.\ W.\, Goss, W.\ M.\, 
Fomalont, E.\ B.\, \&  Benson, J.\ M.\, 2001, \apj, 549
\bibitem[Condon et al. 1998]{condon98} Condon, et al. 1998, AJ, 115, 1693
\bibitem[Cordes \& Helfand 1980]{cordes80} Cordes, J.\ M.\ \& 
Helfand, D.\ J.\ 1980, \apj, 239, 640
\bibitem[Cordes, Pidwerbetsky \& Lovelace 1986]{cpl86}
Cordes, J. M., Pidwerbetsky, A. \& Lovelace, R. V. E. L. 1986,
\apj, 310, 737
\bibitem[Cordes et al. 1993]{cordes93} Cordes, 
J.\ M., Romani, R.\ W., \& Lundgren, S.\ C.\ 1993, \nat, 362, 133 
\bibitem[Cordes \& Chernoff 1997]{cordes97} Cordes, J.\ M., \& Chernoff, D.\ F., 1997, \apj, 482, 971
\bibitem[Cordes \& Chernoff 1998]{cher98} Cordes, J.\ M.\ \& 
Chernoff, D.\ F.\ 1998, \apj, 505, 315 
\bibitem[Cordes \& Rickett 1998]{cordes98} Cordes, J.\ M., \& Rickett, B.\ J. 1998, \apj, 507, 846
\bibitem[Dowd et al. 2000]{dowd00} Dowd, A., Sisk, 
W., \& Hagen, J.\ 2000, ASP Conf.\ Ser.\ 202: IAU Colloq.\ 177: Pulsar 
Astronomy - 2000 and Beyond, 275 
\bibitem[Egger \& Aschenbach 1995]{egger95} Egger, R.\ J.\ \& 
Aschenbach, B.\ 1995, \aap, 294, L25
\bibitem[Fey et al. 1996]{fey96} Fey, A.\ L., 
Clegg, A.\ W., \& Fiedler, R.\ L.\ 1996, \apj, 468, 543 
\bibitem[Gaensler \& Frail 2000]{gaensler2000} Gaensler, B.\ M.\ \& 
Frail, D.\ A.\ 2000, \nat, 406, 158
\bibitem[Gould \& Lyne 1998]{gould98} Gould, D.\ M.\ \& Lyne, 
A.\ G.\ 1998, \mnras, 301, 235 
\bibitem[Hartman et al. 1999]{hart99} Hartman, R.\ C.\ et 
al.\ 1999, \apjs, 123, 79
\bibitem[Haslam et al. 1982]{haslam82} 
Haslam, C.\ G.\ T., Stoffel, H., Salter, C.\ J., \& Wilson, W.\ E.\ 1982, 
\aaps, 47, 1  
\bibitem[Heiles et al. 1980]{heiles80} Heiles, C., Chu, Y.\ -., 
Troland, T.\ H., Reynolds, R.\ J., \& Yegingil, I.\ 1980, \apj, 242, 533
\bibitem[Heiles 1998]{heiles98} Heiles, C.\ 1998, Berlin 
Springer Verlag Lecture Notes in Physics, v.506, 506, 229 
\bibitem[Gott, Gunn, \& Ostriker 1970]{gott1970} Gott, J.\ R.\ 
I., Gunn, J.\ E., \& Ostriker, J.\ P.\ 1970, \apjl, 160, L91 
\bibitem[Lommen et al. 2000]{lommen00} Lommen, A.\ N., Zepka, 
A., Backer, D.\ C., McLaughlin, M., Cordes, J.\ M., Arzoumanian, Z., \& 
Xilouris, K.\ 2000, \apj, 545, 1007 
\bibitem[Lorimer et al. 1995]{lorimer95} 
Lorimer, D.\ R., Yates, J.\ A., Lyne, A.\ G., \& Gould, D.\ M.\ 1995, 
\mnras, 273, 411 
\bibitem[Lorimer et al. 1997]{lor97} Lorimer, D. R., Bailes, M., \& Harrison, P. A. 1997, MNRAS,
289, 592
\bibitem[Lyne \& Lorimer 1994]{lyne94} Lyne, A.\ G.\ \& 
Lorimer, D.\ R.\ 1994, \nat, 369, 127
\bibitem[Manchester 1971]{man71} Manchester, R.\ N.\ 1971, 
\apjs, 23, 283
\bibitem[Manchester \& Johnston 1995]{man95} Manchester, R.\ 
N.\ \& Johnston, S.\ 1995, \apjl, 441, L65 
\bibitem[Manchester et al. 1998]{man98} Manchester, 
R.\ N., Han, J.\ L., \& Qiao, G.\ J.\ 1998, \mnras, 295, 280
\bibitem[McCulloch  et al. 1978]{mcculloch78}
McCulloch, P.\ M., Hamilton, P.\ A., 
Manchester, R.\ N., \& Ables, J.\ G.\ 1978, \mnras, 183, 645 
\bibitem[McLaughlin et al. 1999]{mclaughlin99} McLaughlin, M.\ A., Cordes, J.\ M.,
Hankins, T.\ H., \& Moffett, D.\ A.\ 1999, \apj, 512, 929 
\bibitem[McLaughlin et al. 2000]{mclaughlin00} McLaughlin, M.\ A., Arzoumanian, Z., \& Cordes, J.\ M., 2000, In ASP Conf. Ser. 202, Pulsar Astronomy - 2000 and Beyond, ed. M.\ Kramer, N.\ Wex, \& R. Weilebinksi, (San Fransisco: ASP) pg. 41
\bibitem[McLaughlin \& Cordes 2000]{mclaughlin2000} McLaughlin, M.\ 
A.\ \& Cordes, J.\ M.\ 2000, \apj, 538, 818
\bibitem[Rasio \& Shapiro 1995]{ras95} Rasio, F.\ A.\ \& 
Shapiro, S.\ L.\ 1995, \apj, 438, 887 
\bibitem[Rickett 1990]{ric90} Rickett, B.~J., 1990, \araa, 28, 561
\bibitem[Salter 1983]{salter83} Salter, C.\ J.\ 1983, Bulletin 
of the Astronomical Society of India, 11, 1 
\bibitem[Scheuer 1968]{sch68} Scheuer, P. A.\ G., 1968, Nature, 218, 920
\bibitem[Shklovskii 1970]{sh70} Shklovskii, I.\ S.\ 1970, 
Soviet Astronomy, 13, 562 
\bibitem[Sofue 1977]{sofue77} Sofue, Y.\ 1977, \aap, 60, 327
\bibitem[Taylor \& Weisberg 1989]{tay89} Taylor, J.\ H. \& Weisberg, J.\ M. 1989, \apj, 345, 132
\bibitem[Taylor et al. 1993]{taylor933} Taylor, 
J.\ H., Manchester, R.\ N., \& Lyne, A.\ G.\ 1993, \apjs, 88, 529 
\bibitem[Taylor \& Cordes 1993]{taylor93} Taylor, J.\ H. \& Cordes, J.\ M. 1993,\apj, 411, 674
\bibitem[von Hoensbroech et al. 1998]{von98} 
von Hoensbroech, A., Kijak, J., \& Krawczyk, A.\ 1998, \aap, 334, 571
\bibitem[Weisberg et al. 1999]{weisberg99} Weisberg, J.\ M.\ et 
al.\ 1999, \apjs, 121, 171
\bibitem[Wu et al. 1993]{wu93} Wu, X., 
Manchester, R.\ N., Lyne, A.\ G., \& Qiao, G.\ 1993, \mnras, 261, 630
\bibitem[Xilouris \& Kramer 1996]{xilouris96} Xilouris, K.\ M.\ 
\& Kramer, M.\ 1996, ASP Conf.\ Ser.\ 105: IAU Colloq.\ 160: Pulsars: 
Problems and Progress, 245 
\bibitem[Zepka et al. 1996]{zepka96} 
Zepka, A., Cordes, J.\ M., Wasserman, I., \& Lundgren, S.\ C.\ 1996, \apj, 
456, 305  
\end{thebibliography}
\end{document}